\documentclass{emulateapj}
\usepackage{apjfonts}

\begin{document}

\title{Deep Optical Photometry of Six Fields in the Andromeda 
Galaxy\altaffilmark{1}}

\author{
Thomas M. Brown\altaffilmark{2}, 
Ed Smith\altaffilmark{2}, 
Henry C. Ferguson\altaffilmark{2},
Puragra Guhathakurta\altaffilmark{3}, 
Jason S. Kalirai\altaffilmark{2},
Randy A. Kimble\altaffilmark{4},
Alvio Renzini\altaffilmark{5},
R. Michael Rich\altaffilmark{6}, 
Allen V. Sweigart\altaffilmark{4},
Don A. VandenBerg\altaffilmark{7}
}

\altaffiltext{1}{Based on observations made with the NASA/ESA {\it Hubble
Space Telescope}, obtained at the Space Telescope Science Institute, 
which is operated by the Association of Universities for Research
in Astronomy, Incorporated, under NASA contract NAS5-26555.}

\altaffiltext{2}{Space Telescope Science Institute, 
Baltimore, MD 21218;  tbrown@stsci.edu, edsmith@stsci.edu, ferguson@stsci.edu,
jkalirai@stsci.edu}

\altaffiltext{3}{University of California Observatories / Lick Observatory, 
University of California, Santa Cruz, CA 95064; 
raja@ucolick.org}

\altaffiltext{4}
{NASA Goddard Space Flight Center, Greenbelt, MD
20771; randy.a.kimble@nasa.gov, allen.v.sweigart@nasa.gov}

\altaffiltext{5}{Osservatorio Astronomico, 
I-35122 Padova, Italy; alvio.renzini@oapd.inaf.it}

\altaffiltext{6}{Department of Physics and Astronomy, 
University of California, Los Angeles, CA 90095;
rmr@astro.ucla.edu}

\altaffiltext{7}{Department of Physics and Astronomy, 
University of Victoria, P.O. Box 3055, Victoria, BC, V8W 3P6, Canada; 
vandenbe@uvic.ca}

\submitted{Accepted for publication in The Astrophysics Journal Supplement
Series}

\begin{abstract}

Using the Advanced Camera for Surveys on the {\it Hubble Space
  Telescope}, we have obtained deep optical images reaching well below
the oldest main sequence turnoff in six fields of the Andromeda
Galaxy.  The fields fall at four positions on the southeast minor
axis, one position in the giant stellar stream, and one position on
the northeast major axis.  These data were obtained as part of three
large observing programs designed to probe the star formation history
of the stellar population in various structures of the galaxy.  In
this paper, we present the images, catalogs, and artificial star tests
for these observing programs as a supplement to the analyses published
previously.  These high-level science products are also archived at
the Multimission Archive at the Space Telescope Science Institute.

\end{abstract}


\section{Introduction}

The Andromeda Galaxy (M31) is an ideal laboratory for testing theories
of spiral galaxy formation and evolution.  It appears to be
representative of giant spiral galaxies in the nearby universe (Hammer
et al.\ 2007), and it is the only giant spiral galaxy where we can
accurately measure the complete star formation history from an
external vantage point, using photometry reaching below the oldest
turnoff on the stellar main sequence.  Using the Advanced Camera for
Surveys (ACS) on the {\it Hubble Space Telescope (HST)}, we obtained
deep optical photometry of six fields in M31 (Figure 1), with which we
reconstructed the star formation history at various points in the
spheroid, disk, and giant stellar stream (Brown et al.\ 2003, 2006,
2007, 2008).  As a supplement to those studies, we present in this
paper the reduced images and their associated photometric catalogs,
along with characterizations of the photometric scatter and
completeness derived from artificial star tests.  These high-level
science products (HLSPs) can facilitate comparison with other studies of
stellar populations in nearby galaxies or aid in the planning of
additional observations in Andromeda itself.

\begin{figure}[h]
\epsscale{1.2}
\plotone{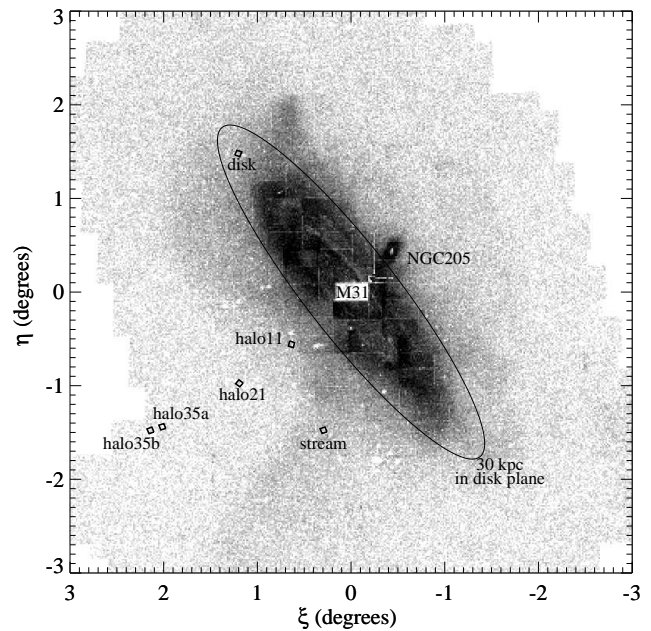}
\epsscale{1.0}
\vskip -0.1in
\caption{The stellar density in the Andromeda vicinity, 
from counts of RGB stars (Ferguson et al.\ 2002).  
Our six fields are indicated ({\it labeled boxes}).  An ellipse
marks the area within 30 kpc of the galactic center in the inclined
disk plane ({\it labeled}).  Later surveys that go wider and deeper
(e.g., Ibata et al. 2007) show an even greater wealth of substructure
in the system.}
\end{figure}

\section{Observations and Data Reduction}

The characteristics of the six fields 
shown in Figure 1 are summarized in Table 1.
Images were obtained using the Wide
Field Camera on the ACS, in the F606W (broad $V$) and F814W
($I$) filters.  These data were obtained as part of three Large ($>$
100 orbits) Guest Observer (GO) programs (GO-9453, GO-10265, and
GO-10816) in {\it HST} Observing Cycles 11, 13, and 15.  

\begin{table*}[t]
\begin{center}
\caption{Characteristics of M31 Fields}
\begin{tabular}{lrccrrl}
\tableline
       & Program & R.A.       & Dec.     & F606W  & F814W  &      \\
Field  & Number & (J2000)    & (J2000)  & (ksec) & (ksec) & Date \\
\tableline
halo11 & GO-9453 & 00:46:07.1 & 40:42:39 & 138.6  & 161.3 & 2 Dec 2002 -- 1 Jan 2003 \\ 
stream & GO-10265 & 00:44:18.2 & 39:47:32 &  52.8  &  78.1 & 30 Aug 2004 -- 4 Oct 2004 \\
disk   & GO-10265 & 00:49:08.6 & 42:45:02 &  52.8  &  78.1 & 11 Dec 2004 -- 18 Jan 2005 \\
halo21 & GO-10816 & 00:49:05.1 & 40:17:32 &  28.7  &  47.8 &  9 Aug 2006 -- 28 Aug 2006 \\
halo35a& GO-10816 & 00:53:28.1 & 39:49:46 &  28.1  &  46.9 & 12 Oct 2006 -- 18 Oct 2006 \\
halo35b& GO-10816 & 00:54:08.5 & 39:47:26 &  28.1  &  51.6 & 18 Oct 2006 --  6 Jan 2007 \\
\tableline
\end{tabular}
\end{center}
\end{table*}

The data properties are not uniform across observing cycles.  The
primary distinction is that each program obtained systematically
shallower images than the preceding one.  When the original program
was planned for the halo11 field, no images of M31 had reached
significantly below the luminosity of the horizontal branch, let alone
the main sequence turnoff.  Given the uncertainties, the original
program was designed to reach $\sim$1.5~mag below the oldest main
sequence turnoff that might be present, in order to ensure an
unambiguous characterization of the detailed star formation history.
The position of the halo11 field on the minor axis was chosen to
optimize the crowding (trading population sample size against
photometric uncertainties).  In addition, the halo11 field was chosen
to be in the vicinity of other historical probes of the M31 spheroid.
From an analysis of subsets of the data from the original program
(GO-9453), we were able to empirically determine how the fidelity of
the recovered star formation history depended upon the depth of the
photometry.  This analysis showed that a depth of $\sim$1~mag below
the oldest main sequence turnoff was sufficient to characterize the
star formation history.  By using this shallower depth in our second
program (GO-10265), we were able to determine the star formation
history in two fields without increasing the overall size of the
observing program.  As with the first program, the locations of the
fields within the targeted structures (disk and giant stellar stream)
were again chosen to provide near-optimal levels of crowding.  In
contrast, the third program (GO-10816) targeted regions of the M31
spheroid far sparser than those in the first two programs.  Because it
was necessary to trade depth against the number of fields to ensure an
adequate sample size, the third program was designed to reach
$\sim$0.8~mag below the oldest main sequence turnoff in each of its
fields.

Besides the above differences in the photometric depth between
programs, there were also differences among the fields of the third
program.  That program planned to obtain one halo field at 21 kpc and
three halo fields at 35 kpc along the minor axis, thereby providing
approximately the same number of stars in each of these two locations.
Due to a hardware failure in the ACS during Cycle 15,
the third program was terminated before it obtained any observations
in the final 35 kpc field, and so the sample size at 35 kpc is
approximately 2/3 of that at 21 kpc.  Furthermore, one of the fields
in the third program (halo35b) had significant deviations from the
planned F814W exposures.  Guide star problems caused some
exposures to fail outright and other exposures to be significantly
offset ($\sim$80$^{\prime\prime}$) from the planned pointing.  The
problematic exposures were rescheduled with a 102$^{\rm o}$ change in
orientation.

Our reduction process for the first two programs 
was fully documented in Brown et al.\ (2006), to
which we refer readers for details.  Note that the catalog produced by
Brown et al.\ (2006) for the halo11, stream, and disk fields
was subject to more rigorous detection thresholds
and culling procedures than that produced by Brown et al.\ (2003) for the
halo11 field; while the
earlier catalog is thus deeper, it also included more spurious sources.
The reduction for the
subsequent fields (halo21, halo35a, and halo35b) was documented in
Brown et al.\ (2007, 2008).  The process used in the later fields was
similar to that in Brown et al.\ (2006), with the additional correction
for charge transfer inefficiency (CTI).  A CTI correction was
unnecessary for the photometry in the first two programs, given the
crowded field, high background, and relative youth of the detector.

We briefly summarize the reduction process here.  The individual
exposures for each bandpass in a given field were registered using the
positions of bright stars well-detected in each exposure.  The
exposures were then combined using the DRIZZLE package (Fruchter \&
Hook 2002), including correction for geometric distortion, rescaling
to 0.03$^{\prime\prime}$ pixel$^{-1}$, and rejection of cosmic rays
and hot pixels.  The individual frames in each field were drizzled to
a 7500$\times$7500 pixel output image, and then this output image was
cropped to a 7000$\times$7350 pixel image with the World Coordinate
System in the header intact.  As part of this process, the sky
background was subtracted from individual frames prior to drizzling
(to avoid increasing the noise in a process that essentially
interlaces the individual frames), and then the sky background was
restored to the coadded images (to ensure that the counting statistics
in subsequent object detection and photometry routines were
appropriate).  PSF-fitting photometry was then performed with the
DAOPHOT-II software of Stetson (1987).  The PSF-fitting photometry was
corrected to agree with aperture photometry of isolated stars, with
the zeropoints calibrated at the 1\% level.  For the fields in the
third program, the CTI correction of Riess \& Mack (2005) was applied.
Our photometry is in the STMAG system: $m= -2.5 \times $~log$_{10}
f_\lambda -21.1$~mag, where $f_\lambda = $ e$^- \times {\rm
  PHOTFLAM/EXPTIME}$, where EXPTIME is the exposure time, and PHOTFLAM
is $7.906 \times 10^{-20}$ erg s$^{-1}$ cm$^{-2}$ \AA$^{-1}$ /
(e$^{-}$ s$^{-1}$) for the F606W filter and $7.072 \times 10^{-20}$
erg s$^{-1}$ cm$^{-2}$ \AA$^{-1}$ / (e$^{-}$ s$^{-1}$) for the F814W
filter.  For those more familiar with the ABMAG system,
ABMAG~=~STMAG~$-0.169$ for $m_{F606W}$, and ABMAG~=~STMAG~$-0.840$ mag
for $m_{F814W}$.  

Note that Brown et al. (2006) shifted the stream photometry 0.03~mag
brighter in $m_{F814W}$ and 0.014~mag redder in $m_{F606W}-m_{F814W}$
color.  These shifts produced excellent agreement between the
distributions of stars on the horizontal branch and red giant branch (RGB)
in the halo11 and stream fields.  The shift in brightness is justified
by the fact that the stream is located behind Andromeda (McConnachie
et al.\ 2003), while the shift in color is well within the calibration
and reddening uncertainties.  The stream catalog presented here does
not include these shifts in color and magnitude.

\begin{table*}[t]
\begin{center}
\caption{Catalog for halo11 field}
\begin{tabular}{rrcccccc}
\tableline
x & y & R.A.& Dec. & $m_{F606W}$ & $m_{F606W}$ error & $m_{F814W}$ & $m_{F814W}$ error \\
(pixels) & (pixels) & (J2000) & (J2000) & (mag) & (mag) & (mag) & (mag) \\
\tableline
  65.11& 310.17 & 00:46:00.99 & +40:40:28.20& 28.98 & 0.04& 29.35 & 0.06 \\
  65.61& 313.96 & 00:46:01.00 & +40:40:28.24& 30.18 & 0.12& 30.20 & 0.09 \\
 132.41& 314.27 & 00:46:00.96 & +40:40:30.18& 31.08 & 0.25& 31.07 & 0.18 \\
 144.49& 315.30 & 00:46:00.95 & +40:40:30.54& 30.43 & 0.13& 30.39 & 0.08 \\
 269.93& 314.81 & 00:46:00.86 & +40:40:34.17& 28.46 & 0.03& 28.92 & 0.03 \\
\tableline
 & & & & & & & \\
\multicolumn{8}{l}{Only a portion of this table is shown here to 
demonstrate its form and content.} \\
\multicolumn{8}{l}{A machine-readable 
version of the full table is available.} \\ 
\end{tabular}
\end{center}
\end{table*}

\begin{table*}[t]
\begin{center}
\caption{Catalog for stream field}
\begin{tabular}{rrcccccc}
\tableline
x & y & R.A.& Dec. & $m_{F606W}$ & $m_{F606W}$ error & $m_{F814W}$ & $m_{F814W}$ error \\
(pixels) & (pixels) & (J2000) & (J2000) & (mag) & (mag) & (mag) & (mag) \\
\tableline
  34.30& 312.88 & 00:44:12.71 & +39:49:46.58& 28.98 & 0.06 & 29.52 & 0.06 \\
 179.08& 318.11 & 00:44:13.06 & +39:49:44.81& 29.21 & 0.07 & 29.65 & 0.08 \\
 171.98& 318.93 & 00:44:13.04 & +39:49:44.87& 28.44 & 0.05 & 28.98 & 0.04 \\
  35.41& 320.12 & 00:44:12.71 & +39:49:46.37& 29.23 & 0.07 & 29.68 & 0.09 \\
 217.24& 320.04 & 00:44:13.15 & +39:49:44.33& 30.33 & 0.21 & 30.88 & 0.25 \\
\tableline
 & & & & & & & \\
\multicolumn{8}{l}{Only a portion of this table is shown here to 
demonstrate its form and content.} \\
\multicolumn{8}{l}{A machine-readable 
version of the full table is available.} \\ 
\end{tabular}
\end{center}
\end{table*}

\begin{table*}[t]
\begin{center}
\caption{Catalog for disk field}
\begin{tabular}{rrcccccc}
\tableline
x & y & R.A.& Dec. & $m_{F606W}$ & $m_{F606W}$ error & $m_{F814W}$ & $m_{F814W}$ error \\
(pixels) & (pixels) & (J2000) & (J2000) & (mag) & (mag) & (mag) & (mag) \\
\tableline
  47.06& 311.27 & 00:49:04.00 & +42:42:41.95& 29.61 & 0.08 & 29.79 & 0.07 \\
 229.83& 314.39 & 00:49:03.82 & +42:42:47.03& 27.92 & 0.03 & 28.46 & 0.03 \\
 134.46& 315.17 & 00:49:03.92 & +42:42:44.41& 29.89 & 0.13 & 30.09 & 0.10 \\
 170.79& 316.31 & 00:49:03.88 & +42:42:45.43& 29.18 & 0.07 & 29.84 & 0.10 \\
  91.33& 318.42 & 00:49:03.97 & +42:42:43.26& 29.12 & 0.07 & 29.66 & 0.08 \\
\tableline
 & & & & & & & \\
\multicolumn{8}{l}{Only a portion of this table is shown here to 
demonstrate its form and content.} \\
\multicolumn{8}{l}{A machine-readable 
version of the full table is available.} \\ 
 & & & & & & & \\
 & & & & & & & \\
\end{tabular}
\end{center}
\end{table*}

The catalogs were then culled to remove problematic stars (artifacts,
blends, extended objects, etc.) using object sharpness, proximity to
brighter stars, and the map of underexposed areas in the image (due to
dithering the image edges and the detector gap).  In the halo11 field,
stars within 500 pixels (15$^{\prime\prime}$) of the globular cluster
GC312 were also removed from the catalog.  This radius was chosen to
safely exceed the cluster's tidal radius of 10$^{\prime\prime}$
(Holland et al. 1997); the deep photometry of GC312 was analyzed
separately by Brown et al.\ (2004).  Given the problems with the 35
kpc sample (one field missing outright and another field with pointing
irregularities in the F814W band), an exception was made to the
culling process for the halo35b field to maximize the number of stars
in the 35 kpc sample.  Instead of culling any stars falling in a part
of the image with off-nominal exposure time, we included regions where
the exposure depth was within 0.2~mag of nominal; this includes
regions of the image that are 0.1~mag deeper than planned
(approximately 50\% of the image area) and regions that are 0.2~mag
shallower than planned (approximately 10\% of the image area).  Even
though much of the halo35b F814W image is deeper than planned, the sky
background for the field is also higher than that in the halo35a
field, because it executed later in the year, when the Sun angle for
Andromeda was smaller.

To characterize the scatter and incompleteness in the photometric
catalogs, thousands of artificial star tests were performed on the
images from each field; each test inserted and blindly recovered
hundreds to thousands of stars, but the total number of artificial
stars tested in each field exceeded 4 million.  The artificial star
tests were uniformly distributed over the regions of each image
spanned by its associated catalog.  For the halo35b field, the stars
were inserted and recovered in a manner consistent with the varying
exposure depth in the image (i.e., with stellar magnitudes scaled to
the exposure time of the region where the stars were inserted).

\section{High-Level Science Products}

The HLSPs include coadded images, masks, catalogs, and artificial star
tests.  All of these products have been been delivered to the
Multimission Archive at the Space Telescope Science Institute (MAST).
The catalogs are also included as electronic tables here (Tables 2--7)
with their own naming convention specific to the Journal.  The HLSPs
follow the MAST naming convention.  Each filename begins with {\tt
  hlsp\_andromeda\_hst\_acs-wfc\_}, concatenating {\tt hlsp} with the
project name, mission name, and instrument name.  The remainder of the
filename concatenates the field name (given in Table 1), the relevant
filter(s), the version number, the product type, and a suffix.  The
version number for the products discussed here is {\tt v2}, because
the process we developed was changed significantly between Brown et
al.\ (2003) and Brown et al.\ (2006).  The product types are images
({\tt img}), masks ({\tt msk}), catalogs ({\tt cat}), and artificial
star tests ({\tt art}).  The catalogs are ascii machine-readable
tables with suffix {\tt txt}, while the remaining products are
FITS files with suffix {\tt fits}.  All of the FITS files are in
standard FITS format without multiple FITS extensions.

\begin{table*}[t]
\begin{center}
\caption{Catalog for halo21 field}
\begin{tabular}{rrcccccc}
\tableline
x & y & R.A.& Dec. & $m_{F606W}$ & $m_{F606W}$ error & $m_{F814W}$ & $m_{F814W}$ error \\
(pixels) & (pixels) & (J2000) & (J2000) & (mag) & (mag) & (mag) & (mag) \\
\tableline
 280.50& 233.02 & 00:49:06.77 & +40:19:57.06& 29.89 & 0.17 & 30.18 & 0.19 \\
 765.69& 249.27 & 00:49:07.52 & +40:19:45.29& 27.65 & 0.04 & 28.02 & 0.03 \\
 528.11& 253.14 & 00:49:07.13 & +40:19:50.83& 29.53 & 0.11 & 30.14 & 0.17 \\
 379.68& 253.21 & 00:49:06.89 & +40:19:54.34& 29.08 & 0.07 & 29.54 & 0.07 \\
 623.00& 253.86 & 00:49:07.28 & +40:19:48.58& 29.62 & 0.15 & 30.04 & 0.11 \\
\tableline
 & & & & & & & \\
\multicolumn{8}{l}{Only a portion of this table is shown here to 
demonstrate its form and content.} \\
\multicolumn{8}{l}{A machine-readable 
version of the full table is available.} \\ 
\end{tabular}
\end{center}
\end{table*}

\begin{table*}[t]
\begin{center}
\caption{Catalog for halo35a field}
\begin{tabular}{rrcccccc}
\tableline
x & y & R.A.& Dec. & $m_{F606W}$ & $m_{F606W}$ error & $m_{F814W}$ & $m_{F814W}$ error \\
(pixels) & (pixels) & (J2000) & (J2000) & (mag) & (mag) & (mag) & (mag) \\
\tableline
 427.99& 229.71 & 00:53:17.44 & +39:50:58.61& 28.91 & 0.08 & 29.48 & 0.10 \\
 145.40& 230.58 & 00:53:16.74 & +39:50:56.24& 29.17 & 0.10 & 29.65 & 0.11 \\
1214.17& 237.96 & 00:53:19.42 & +39:51:04.88& 29.31 & 0.10 & 29.92 & 0.14 \\
1049.68& 239.57 & 00:53:19.01 & +39:51:03.47& 29.23 & 0.10 & 29.89 & 0.11 \\
1538.71& 245.01 & 00:53:20.24 & +39:51:07.37& 29.12 & 0.10 & 29.52 & 0.10 \\
\tableline
 & & & & & & & \\
\multicolumn{8}{l}{Only a portion of this table is shown here to 
demonstrate its form and content.} \\
\multicolumn{8}{l}{A machine-readable 
version of the full table is available.} \\ 
\end{tabular}
\end{center}
\end{table*}

\begin{table*}[t]
\begin{center}
\caption{Catalog for halo35b field}
\begin{tabular}{rrcccccc}
\tableline
x & y & R.A.& Dec. & $m_{F606W}$ & $m_{F606W}$ error & $m_{F814W}$ & $m_{F814W}$ error \\
(pixels) & (pixels) & (J2000) & (J2000) & (mag) & (mag) & (mag) & (mag) \\
\tableline
  50.27& 221.81 & 00:53:56.66 & +39:48:30.44& 28.46 & 0.06 & 28.83 & 0.06 \\
 906.23& 228.29 & 00:53:58.78 & +39:48:38.32& 28.53 & 0.06 & 28.75 & 0.05 \\
 509.27& 228.74 & 00:53:57.80 & +39:48:34.57& 28.38 & 0.06 & 28.68 & 0.06 \\
 502.70& 244.08 & 00:53:57.80 & +39:48:34.07& 29.98 & 0.16 & 29.55 & 0.08 \\
1567.05& 250.23 & 00:54:00.43 & +39:48:43.92& 28.91 & 0.07 & 29.43 & 0.10 \\
\tableline
 & & & & & & & \\
\multicolumn{8}{l}{Only a portion of this table is shown here to 
demonstrate its form and content.} \\
\multicolumn{8}{l}{A machine-readable 
version of the full table is available.} \\ 
 & & & & & & & \\
 & & & & & & & \\
\end{tabular}
\end{center}
\end{table*}

Each field has two coadded images -- one for each band (F606W and
F814W).  The floating-point array in each file is in units of counts,
with all parts of each image scaled to a common exposure time (given
by the EXPTIME keyword in the image header and in Table 1).  The
geometric distortion in the ACS is significant, such that there is an
irregular border of unexposed pixels in each rectangular image after
correction for geometric distortion.  This border has been set to a
value of 3$\times$10$^7$ counts so that it can be easily masked in
image analysis software.  There are also a handful of negative pixels
near this border -- an artifact of the DRIZZLE process in places where
the image is severely underexposed.

Associated with each field is a mask file.  The byte
array in each mask file is set to 0 in regions of the image valid for
the catalog and artificial star tests, and set to 1 in regions that
are invalid.  The masking in these files is much more aggressive than
that given by the pixels set to 3$\times$10$^7$ counts in the
coadded images described above, since the latter is merely intended to
flag pixels with little or no exposure time.  Here, the mask is set to 0
where the exposure time deviates significantly from the planned exposure
time due to dithers.

The catalogs for each field have 8 columns: x (pixels, 1-indexed), y
(pixels, 1-indexed), right ascension (J2000), declination (J2000),
$m_{F606W}$ (magnitudes), $m_{F606W}$ error (magnitudes), $m_{F814W}$
(magnitudes), and $m_{F814W}$ error (magnitudes).  
The error reported in the catalogs is simply that given by DAOPHOT-II.
We do not use these errors in our own analyses; instead, when
comparing models to the data, we scatter our models according to the
results of artificial star tests.  The right ascension and declination
for each star comes from the World Coordinate System specified in the
image headers, and is only as accurate as the coordinates of the guide
stars used in the observations.  The first two programs executed with
the first version of the Guide Star Catalog, with an associated
astrometric uncertainty of $\sim$1$^{\prime\prime}$; the third program executed
with the second version of the Guide Star Catalog, with an associated
astrometric uncertainty of $\sim$0.3$^{\prime\prime}$.  Truncated
versions of the catalog for each field are provided in Tables 2--7.
The full catalog for each field is provided as a machine-readable
electronic table in the online edition of The Astrophysical
Journal. Binned versions of all six catalogs are shown in Figure 2.

\begin{figure*}[t]
\epsscale{1.1}
\plotone{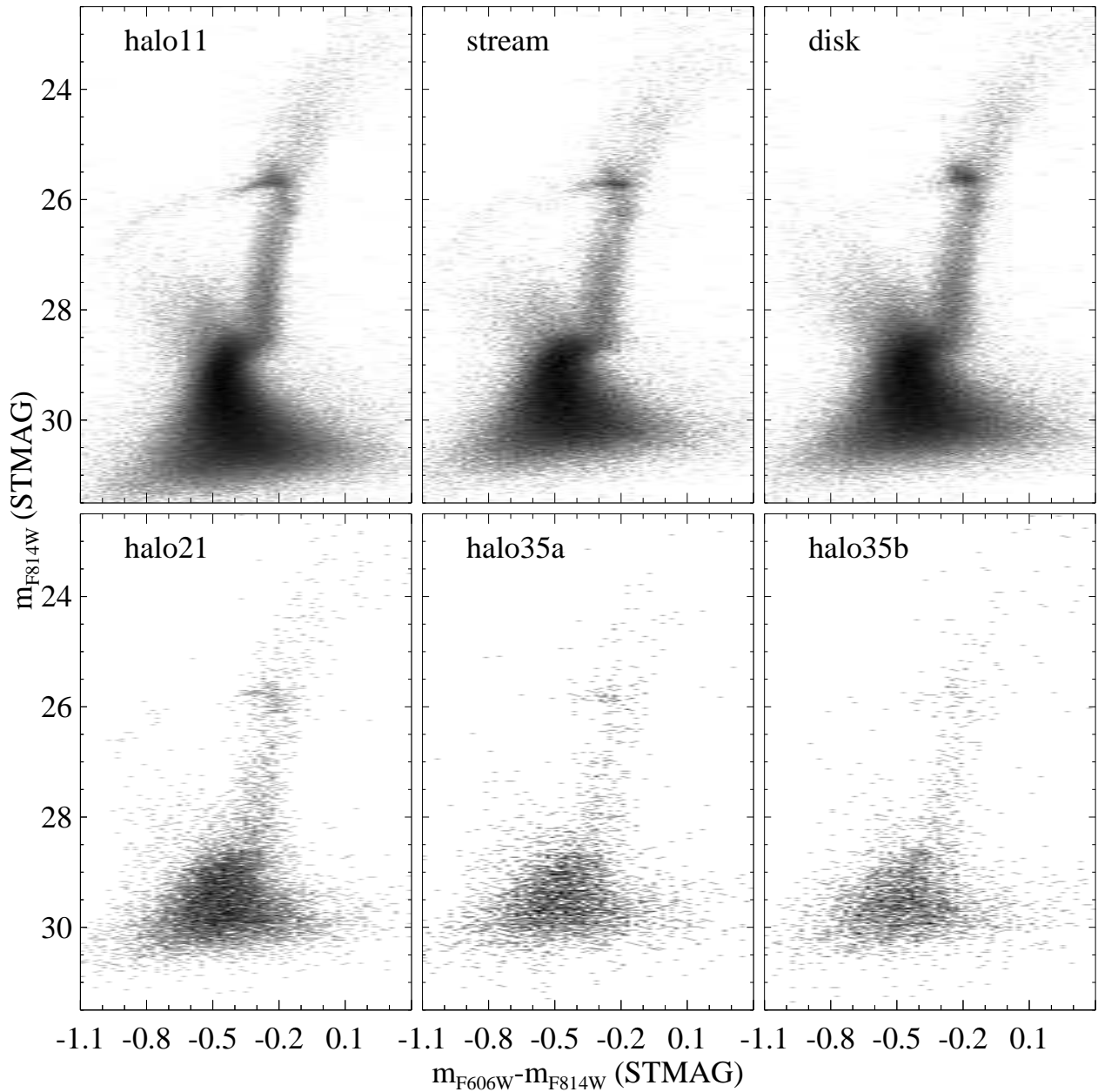}
\epsscale{1.0}
\caption{The binned color-magnitude diagrams for the six fields
given in Table 1, shown at a logarithmic stretch.}
\end{figure*}

The artificial star tests are provided as 4-dimensional floating-point
arrays.  These arrays give the scattering
kernel at each point in the color-magnitude diagram (CMD).  Examples
of these scattering kernels are given in Figure 3.  The first
dimension of the array is the input $m_{F606W}-m_{F814W}$ color,
binned from $-1.1$~mag to 0.4~mag with 16 steps of 0.1~mag.  The
second dimension of the array is the input $m_{F814W}$ magnitude,
binned from 31.5~mag to 22.5~mag with 91 steps of 0.1~mag.  The third
dimension of the array is the output $m_{F606W}-m_{F814W}$ color
relative to the input color, binned from $-0.51$~mag to 0.51~mag with
51 steps of 0.02~mag.  The fourth dimension of the array is the output
$m_{F814W}$ magnitude relative to the input magnitude, binned from
$0.51$~mag to $-0.51$~mag with 51 steps of 0.02~mag.  To avoid
confusion, it is worth stressing the sign convention in the second and
fourth dimensions reflects that in the astronomical magnitude system.
The output scattering kernel (i.e., the third and fourth dimensions of
the array) is normalized such that the sum of the elements gives the
completeness.  The scattering kernel is slowly varying on the scale of
the input binning (0.1~mag), while the binning of the scattering
kernel itself (0.02~mag) matches the scale used to compare data and
models in our star formation history analyses (Brown et al.\ 2006,
2007, 2008).

For example, using a 1-indexed system, the elements in the artificial
star test arrays that correspond to (5,21,*,*) give the scattering
kernel for artificial stars input at colors within 0.05~mag of
$m_{F606W}-m_{F814}$~=~$-0.7$~mag and magnitudes within 0.05~mag of
$m_{F814W}$~=~29.5~mag.  The sum of these elements in the halo11 field
is 0.895, implying that the completeness for stars at this color and
magnitude is 89.5\%.  This scattering kernel is displayed in the lower
left-hand panel of Figure 3.

\section{Summary}

In this paper, we have made public the HLSPs associated with three
{\it HST} Large GO programs targeting various structures in the giant
spiral galaxy M31.  The deep images and their associated masks,
catalogs, and artificial star tests have been delivered to the MAST
archive, at the same site used to store HLPSs from {\it HST} Treasury
programs.  The catalogs are also available as electronic tables in
this paper.  The information provided should be sufficient to make
detailed comparisons between the data in our observing programs and
data in other programs targeting nearby stellar populations.

\acknowledgements

Support for Programs GO-9453, GO-10265, and GO-10816 was provided by
NASA through a grant from the Space Telescope Science Institute, which
is operated by the Association of Universities for Research in
Astronomy, Incorporated, under NASA contract NAS5-26555.  P. Royle and
S. Meyett were enormously helpful in the scheduling and execution of
these Large {\it HST} programs.  PG and JSK would like to acknowledge
support from NASA grants associated with these {\it HST} programs and from
the following NSF grants to UCSC: AST-0307966, AST-0507483, and
AST-0607852.  RMR would like to acknowledge support from grants
associated with these {\it HST} programs and also NSF-AST-03-07931.  We are
grateful to P.\ Stetson for his DAOPHOT code.

\begin{figure*}[t]
\epsscale{1.1}
\plotone{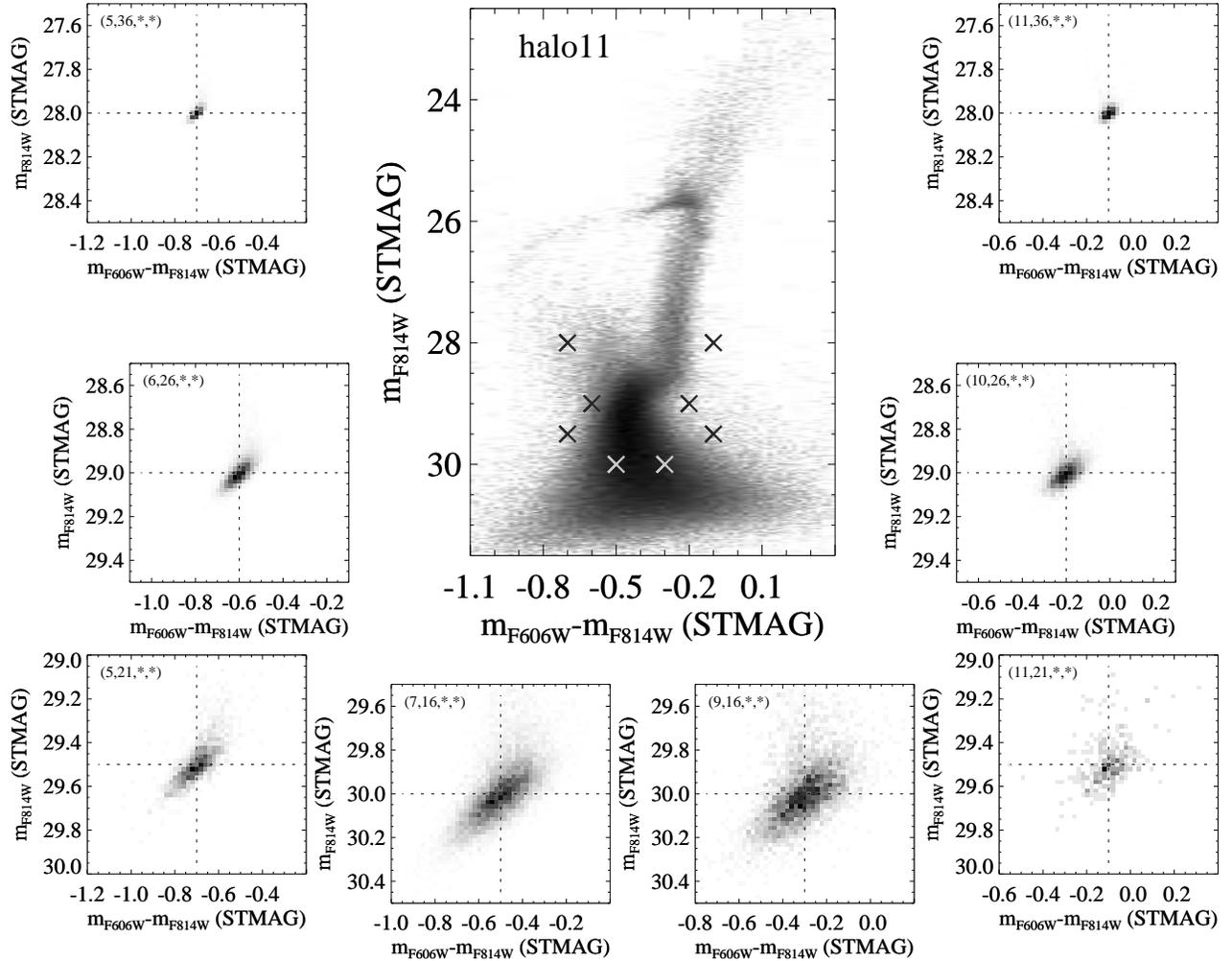}
\epsscale{1.0}
\caption{
The binned CMD for the halo11 field ({\it large central panel}), with
the scattering kernels ({\it small panels}) corresponding to 8
locations in the CMD ({\it grey crosses}).  Each scattering kernel is
labeled with the corresponding indices in the artificial star test
arrays.  The CMD and scattering kernels are shown at a logarithmic
stretch.}
\end{figure*}


\begin{references}

\reference{} %
Brown, T.M., et al.\ 2007, ApJ, 658, L95

\reference{} %
Brown, T.M., et al.\ 2008, ApJ, 685, L121

\reference{} %
Brown, T.M., Ferguson, H.C., Smith, E., Kimble, R.A., Sweigart, A.V.,
Renzini, A., Rich, R.M., \& VandenBerg, D.A. 2003, ApJ,592, L17

\reference{}
Brown, T.M., Ferguson, H.C., Smith, E., Kimble, R.A., Sweigart, A.V.,
Renzini, A., Rich, R.M., \& VandenBerg, D.A. 2004, ApJ, 613, L125

\reference{} %
Brown, T.M., Smith, E., Ferguson, H.C., Rich, R.M., Guhathakurta, P., 
Renzini, A., Sweigart, A.V., \& Kimble, R.A. 2006, ApJ, 652, 323

\reference{} %
Ferguson, A.M.N., Irwin, M.J., Ibata, R.A., Lewis, G.F., \& Tanvir, N.R. 2002,
AJ, 124, 1452

\reference{} %
Fruchter, A.S., \& Hook, R.N. 2002, PASP, 114, 144

\reference{} %
Hammer, F., Puech, M., Chemin, L., Flores, H., \& Lehnert, M.D. 2007,
ApJ, 662, 322

\reference{}
Holland, S., Fahlman, G.G., \& Richer, H.B. 1997, AJ, 114, 1488

\reference{} %
Ibata, R., Martin, N.F., Irwin, M., Chapman, S., Ferguson, A.M.N.,
Lewis, G.F., \& McConnachie, A.W. ApJ, 671, 1591

\reference{}
McConnachie, A.W., Irwin, M.J., Ibata, R.A., Ferguson, A.M.N., Lewis, G.F.,
\& Tanvir, N. 2003, MNRAS, 343, 1335

\reference{}
Riess, A., \& Mack, J. 2005, in Instrument Science Report ACS 2004-006

\reference{} %
Stetson, P. 1987, PASP, 99, 191

\end{references}
\end{document}